\documentclass[prd,amsmath,amssymb,floatfix,superscriptaddress,notitlepage,nofootinbib,preprintnumbers]{revtex4-1}
\usepackage{amsfonts,amssymb,amsmath,graphicx,color,bm,enumitem}
\definecolor{ultramarine}{rgb}{0.07, 0.04, 0.56}
\definecolor{cadmiumgreen}{rgb}{0.0, 0.42, 0.24}
\definecolor{indigo(dye)}{rgb}{0.0, 0.25, 0.42}
\usepackage[linktocpage=true]{hyperref}
\hypersetup{
colorlinks=true,
citecolor=ultramarine,
linkcolor=cadmiumgreen,
urlcolor=indigo(dye),
}

\newcommand{\f}[2]{\frac{#1}{#2}}  
\newcommand{\mk}[1]{\left( #1 \right)}  
\newcommand{\kk}[1]{\left[ #1 \right]}  
  
\newcommand{\be}{\begin{equation}}
\newcommand{\ee}{\end{equation}}
\newcommand{\Mpl}{M_{\rm Pl}}
\newcommand{\E}{\mathcal{E}}
\renewcommand{\L}{\mathcal{L}}
\newcommand{\EC}{\E_C}
\newcommand{\ED}{\E_D}
\newcommand{\ELC}{\E_{C,\Lambda}}
\newcommand{\ELD}{\E_{D,\Lambda}}
\newcommand{\pa}{\partial}

\begin{document}

\preprint{YITP-19-01}

\title{
Exact black hole solutions in shift-symmetric quadratic degenerate higher-order scalar-tensor theories
}

\author{Hayato Motohashi}
\affiliation{Center for Gravitational Physics, Yukawa Institute for Theoretical Physics, Kyoto University,\ Kyoto 606-8502, Japan}

\author{Masato Minamitsuji}
\affiliation{Centro de Astrof\'{\i}sica e Gravita\c c\~ao  - CENTRA, Departamento de F\'{\i}sica, Instituto Superior T\'ecnico - IST,
Universidade de Lisboa - UL, Av. Rovisco Pais 1, 1049-001 Lisboa, Portugal}

\begin{abstract}
We find various exact black hole solutions in the shift-symmetric subclass of the quadratic degenerate higher-order scalar-tensor (DHOST) theories with linearly time dependent scalar field whose kinetic term is constant. The exact solutions are the Schwarzschild and Schwarzschild-(anti-)de Sitter solutions, and the Schwarzschild-type solution with a deficit solid angle, which are accompanied by nontrivial scalar field regular at the black hole event horizon. We derive the conditions for the coupling functions in the DHOST Lagrangian that allow the exact solutions, clarify their compatibility with the degeneracy conditions, and provide general form of coupling functions as well as simple models that satisfy the conditions.
\end{abstract}

\maketitle

\section{Introduction}
\label{sec1}

Scalar-tensor theories are simple models of modification of gravity with additional degrees of freedom. 
They have served models for explaining the phases of accelerated expansion of the Universe at the early time and/or present time, 
as well as for testing gravity in the strong field regime such as the vicinity of black holes.  
They also have led us to a deeper understanding of gravitation.  
In particular, theories with nonminimal coupling to gravity and/or matter and higher-derivatives in the Lagrangian have 
provided a variety of indications for phenomenology and cosmology.

The quadratic degenerate higher-order scalar-tensor (DHOST) theories~\cite{Langlois:2015cwa} provide a general framework of single-field scalar-tensor theories which contains quadratic order terms of second-order derivatives of the scalar field and allow higher-order Euler-Lagrange equations, but possess 3 degrees of freedom without Ostrogradsky ghosts~\cite{Woodard:2015zca}.
In \cite{Langlois:2015cwa}, the construction of degenerate theory for a toy model of scalar-tensor Lagrangians quadratic to second-order derivatives was clarified in the framework of analytical mechanics, and then was applied to the construction of the Lagrangian of the quadratic DHOST theories.
In the framework of analytical mechanics, the degeneracy conditions 
were systematically derived for the general Lagrangian involving up to second-order derivatives~\cite{Motohashi:2016ftl}, for a toy model involving third-order derivatives~\cite{Motohashi:2017eya}, and finally, for the general Lagrangian involving arbitrary higher-order derivatives~\cite{Motohashi:2018pxg}, which is a completion of an early work~\cite{Motohashi:2014opa} which clarified the first and second degeneracy conditions for a Lagrangian involving arbitrary higher-order derivatives.
The derivation of the degeneracy conditions respects the original spirit of the analytical mechanics, as they are based on ``solely algebraic operations subject to a regular and uniform procedure''~\cite{Lagrange1997}, and hence, applied to construction of a wide class of field theories~\cite{Langlois:2015cwa,Crisostomi:2016czh,Kimura:2016rzw,BenAchour:2016fzp,Crisostomi:2017aim,Crisostomi:2017ugk,Kimura:2018sfs,Naruko:2018akp}.

Having the direct detections of gravitational waves from binary black hole and neutron star mergers~\cite{TheLIGOScientific:2017qsa,GBM:2017lvd,Monitor:2017mdv} and 
the forthcoming multimessenger astronomy and astrophysics in the near future,
application of higher-derivative scalar-tensor theories to black hole solutions are of great interest. 
Black hole solutions with modified profiles in the metric and scalar field from general relativity (GR) have been of prior interests.
On the other hand, if it is the case that the future observation will not detect any deviation from GR prediction,
it is also important to explore solutions same as GR.
Depending on the profile of the scalar field, these solutions can be classified into several classes (see Table~\ref{table1}). 
The simplest case is the constant profile.
In general, theories with arbitrary higher-order derivatives allow GR solutions with constant profile of the scalar field(s) as a solution of the equations of motion 
if the coupling functions satisfy a certain set of regularity conditions~\cite{Motohashi:2018wdq} (see also references therein).

\begin{table}[t]
       \caption{Exact black hole solutions in scalar-tensor theories focusing on those with GR(-like) metrics.
       ``Sch'', ``Sch-a'', ``SdS'', ``S(A)dS'', respectively, denote the Schwarzschild solution, Schwarzschild-type solution with a solid deficit angle, Schwarzschild-de Sitter solution, Schwarzschild-(anti-)de Sitter solution.
       The analyses of \cite{Motohashi:2018wdq,Minamitsuji:2018vuw} include black hole solutions as a special case.
       }
       \begin{tabular}{cccc}
              \hline\hline
              & $g_{\mu\nu}$ & $\phi$ & $\L$ \\ \hline
              \cite{Motohashi:2018wdq} & Any GR solution &  & Multi-scalar-tensor theories with \\ 
              & $G_{\mu\nu}=8\pi GT_{\mu\nu}-\Lambda g_{\mu\nu}$ & $\phi=\text{const.}$ & ~ arbitrary higher-order derivatives ~ \\ 
              &&& in $D$-dimensional spacetime \\ \hline
              \cite{Babichev:2017guv} & Sch(-a) & $\phi(r)$ & Shift-sym. GLPV \\ 
              & (stealth)  & $X=\text{const.}$ &  \\ \hline
              \cite{Minamitsuji:2018vuw} & Vacuum GR solution $R_{\mu\nu}=0$ & $\phi(r)$ & Horndeski subclass where $c_t=c$  \\ 
              &  (stealth) & &(shift sym.\ broken) \\ \hline
              \cite{Babichev:2013cya,Kobayashi:2014eva} & Sch \& S(A)dS & $\phi(t,r)=qt+\psi(r)$ &  Shift-sym.\ Horndeski \\ 
              & (stealth \& self-tuned) & $X=\text{const.}$ &  \\ \hline
              \cite{Babichev:2016kdt,Babichev:2017lmw} & SdS & $\phi(t,r)=qt+\psi(r)$ & Shift-sym.\ GLPV  \\ 
              & (self-tuned) & $X=\text{const.}$ &  
              \\ \hline
              \cite{BenAchour:2018dap} & Sch \& S(A)dS & $\phi(t,r)=qt+\psi(r)$ & Quadratic DHOST \\
              & (stealth \& self-tuned) & $X=-q^2$ & with $A_1=A_2=0$ \\ \hline
              This work & Sch(-a) \& S(A)dS & $\phi(t,r)=qt+\psi(r)$ & Shift-sym.\ quadratic DHOST \\
              & (stealth \& self-tuned) & $X=\text{const.}$ &  
              \\
              \hline\hline
       \end{tabular}
       \label{table1}
\end{table}

Besides GR black hole solutions with the constant profile of the scalar field, 
higher-derivative scalar-tensor theories also 
allow special types of black hole solutions which are absent in GR.
For the classification of black hole solutions, a no-hair theorem plays an important role.
For shift-symmetric scalar-tensor theories with the regularity assumptions,
as shown in~\cite{Hui:2012qt} for the shift-symmetric subclass of Horndeski theory~\cite{Horndeski:1974wa} 
and later generalized in~\cite{Babichev:2017guv} for the shift-symmetric subclass of Gleyzes-Langlois-Piazza-Vernizzi (GLPV) theory~\cite{Gleyzes:2014dya}, 
the static, spherically symmetric, and asymptotically flat black hole solutions with static scalar field must have the Schwarzschild metric and the constant scalar field.
The presence of the canonical kinetic term $-X/2$ with $X:=g^{\mu\nu}\partial_\mu\phi\partial_\nu\phi$ which dominates the property of the scalar field $\phi$ at the asymptotic infinity 
is necessary as an additional assumption of the theorem. 
A violation of some of the assumptions of the no-hair theorem allows one to have hairy black hole solutions.
For instance, by violating the regularity assumptions of the coupling functions of the model, asymptotically flat
hairy black hole solutions were found for the shift-symmetric Horndeski theory~\cite{Sotiriou:2013qea,Sotiriou:2014pfa} and for the shift-symmetric GLPV theory~\cite{Babichev:2017guv}.
These solutions possess metrics different from GR and the static profiles of the scalar field $\phi=\phi(r)$ with general nonconstant $X$,
where $r$ is the radial coordinate.

Focusing on solutions with metrics the same as in GR, there are several types of interesting solutions.
The stealth Schwarzschild solutions were found in~\cite{Babichev:2013cya,Kobayashi:2014eva} 
for the shift-symmetric subclasses of the Horndeski theory
for the linearly time dependent scalar field profile $\phi(t,r)=qt+\psi(r)$
with the assumption of a constant kinetic term $X=\text{const}$.
The term ``stealth'' means that
the metric is independent of the model parameters in the scalar field sector of the Lagrangian
and the existence of the scalar field is hidden in the spacetime geometry.
The time dependent scalar field circumvents the assumption of the no-hair theorem
that $\phi=\phi(r)$.

In the subclass of the shift-symmetric GLPV theories 
which does not have the canonical kinetic term to circumvent the no-hair theorem for the shift-symmetric GLPV theories,
the Schwarzschild-type solution with a deficit solid angle was found in \cite{Babichev:2017guv},
similar to the global monopole solution~\cite{Barriola:1989hx}.
The scalar field has the static profile $\phi=\phi(r)$ and constant $X$.
If the coupling functions satisfy the certain tuned relation,
the solid deficit angle is absent and this solution reduces to the stealth Schwarzschild solution.

Furthermore, recently two types of stealth Schwarzschild solutions were
found in the shift symmetry breaking Horndeski subclass up to the $L_3$ Lagrangian in which gravitational waves propagate with the speed of light~\cite{Minamitsuji:2018vuw} (see also \S\ref{sec3d}).
This solution has the static scalar field profile $\phi=\phi(r)$ and allows nonconstant $X$. 
The breakdown of shift symmetry 
also allows us to circumvent the no-hair  theorem for the shift-symmetric theories.  
It was also shown in~\cite{Minamitsuji:2018vuw} that in the shift symmetric breaking Horndeski theory, the stealth solution does not exist with a more general scalar profile such as $\phi=\phi(t,r)$.

Regarding the case of nonasymptotically flat solutions with a GR metric,
an interesting solution is the self-tuned Schwarzschild-de Sitter solution.
For this type of solution, the metric contains a cosmological constant 
which is independent of a bare cosmological constant that appeared in the action,
and hence, the spacetime metric may not be affected by a sudden change
of the bare cosmological constant or vacuum energy via a phase transition~\cite{Charmousis:2011bf,Charmousis:2011ea,Martin-Moruno:2015bda}.
The self-tuned Schwarzschild-de Sitter solutions were found in the shift-symmetric Horndeski theory~\cite{Babichev:2013cya,Kobayashi:2014eva}, and 
in the shift-symmetric GLPV theory~\cite{Babichev:2016kdt}
and the subclass of it where the propagation speed of gravitational waves $c_t$ coincide with the speed of light $c$~\cite{Babichev:2017lmw}.
For these solutions, $\phi(t,r)=qt+\psi(r)$ and $X=\text{const.}$ were assumed.

In this paper, 
we will clarify the conditions for the existence of the Schwarzschild and 
Schwarzschild-(anti-)de Sitter black hole solutions
under the assumptions of a linearly time dependent profile of the scalar field 
$\phi(t,r)=qt+\psi(r)$ and a constant value of $X=X_0$,
and find concrete models that realize the stealth and self-tuned black hole solutions. 
Our subject would have overlap with the recent work~\cite{BenAchour:2018dap},
which also addressed the conditions for the existence of the stealth and self-tuned solutions 
with $X=-q^2$ in the subclass of the quadratic DHOST theories
with a condition $A_1(\phi,X)=A_2(\phi,X)=0$ for the coupling functions [see \eqref{qdaction} for the notation].
Relativistic star solutions in shift-symmetric quadratic DHOST theories with $c_t=c$ condition were also investigated in Ref.~\cite{Kobayashi:2018xvr}.
Our analysis will focus on exact black hole solutions with $X=\text{cosnt.}$ in the shift-symmetric quadratic DHOST theories
but will not be restricted to the solutions with $X=-q^2$ nor
the models with $A_1(\phi,X)=A_2(\phi,X)=0$.
We will also present the detailed derivation process of these conditions,
and explicit models and solutions.

The rest of the paper is organized as follows.
In \S\ref{sec2}, we define the model and describe our strategy to construct exact black hole solutions.
We assume that the scalar field profile takes the form of $\phi(t,r)=qt+\psi(r)$ and the canonical kinetic term remains constant.
In \S\ref{sec3a} and \S\ref{sec3b}, we apply the strategy to the Schwarzschild and Schwarzschild-(anti-)de Sitter solutions.
As we will see later, we will not need to impose the degeneracy conditions at this stage, 
since they are not directly related to the conditions for the existence of particular solutions~\cite{Motohashi:2018wdq}.
In \S\ref{sec3deg}, we take into account the degeneracy conditions, and investigate all the classes of the quadratic DHOST theories.  We clarify the compatibility between the degeneracy conditions and the conditions to allow the exact solutions obtained in this paper.
In \S\ref{sec3c}, we provide general form of coupling functions of the DHOST theories that satisfy the conditions, which is supplemented by additional conditions in \S\ref{sec3d} for the subclass in which the speed of gravitational waves $c_t$ coincides with the speed of light $c$ and there is no decay of gravitons into dark energy.
In \S\ref{sec4}, we focus on several subclasses of our model and clarify the form of conditions.
In \S\ref{sec5}, we discuss applications of our conditions and find simple models for the constant scalar solution, stealth Schwarzschild solution, and self-tuned Schwarzschild-(anti-)de Sitter solution.
\S\ref{sec6} is devoted to conclusion.

\section{The model}
\label{sec2}

We consider the quadratic DHOST theories defined by~\cite{Langlois:2015cwa}
\be \label{qdaction} S = \int d^4x \sqrt{-g} \kk{ F_0(\phi,X) + F_1(\phi,X) \Box\phi + F_2(\phi,X) R + \sum_{I=1}^5 A_I(\phi,X)L_I^{(2)}},  \ee
where 
$g_{\mu\nu}$ is the metric, $g := {\rm det}(g_{\mu\nu})$, $R$ is the Ricci scalar associated with $g_{\mu\nu}$,
$\phi$ is the scalar field,
$X:= \phi_\mu\phi^\mu$,
\be L_1^{(2)} = \phi^{\mu\nu}\phi_{\mu\nu}, \quad
L_2^{(2)} = (\Box\phi)^2, \quad
L_3^{(2)} = \phi^\mu \phi_{\mu\nu} \phi^\nu \Box\phi , \quad
L_4^{(2)} = \phi^\mu \phi_{\mu\nu} \phi^{\nu\lambda} \phi_\lambda , \quad
L_5^{(2)} = (\phi^\mu \phi_{\mu\nu} \phi^\nu)^2, \ee
with 
$\phi_\mu:= \nabla_\mu\phi$, $\phi_{\mu\nu} := \nabla_\nu\nabla_\mu\phi$,
$\Box\phi:= g^{\mu\nu} \phi_{\mu\nu}$,
and 
$F_i(\phi,X)$ ($i=0,1,2$) and $A_I(\phi,X)$ ($I=1,\cdots, 5$)
are arbitrary functions of $\phi$ and $X$.

The action~\eqref{qdaction} of the quadratic DHOST theories can be also written in the form 
\be \label{acC2} S = \int d^4x \sqrt{-g} \kk{ F_0(\phi,X) + F_2(\phi,X) R + \phi_{\mu\nu} C_2^{\mu\nu} } , \ee
where
\be C_2^{\mu\nu} 
= F_1 g^{\mu\nu} 
+ A_1 \phi^{\mu\nu} 
+ A_2 g^{\mu\nu} \Box \phi 
+ A_3 \phi^\mu\phi^\nu \Box\phi 
+ A_4 \phi^\mu\phi^{\nu\lambda}\phi_\lambda 
+ A_5 \phi^\mu\phi^\nu \phi^\alpha\phi_{\alpha\beta}\phi^\beta . \ee
It is also straightforward to rewrite the action of the cubic-order DHOST theories~\cite{BenAchour:2016fzp} in the form~\eqref{acC2}.
The model~\eqref{acC2} is a subclass of a wide class of higher-derivative theories considered in \cite{Motohashi:2018wdq}.  
In \cite{Motohashi:2018wdq}, the condition for such theories to allow any metric solution satisfying the Einstein equation in 
GR with cosmological constant $\Lambda$
\be G_{\mu\nu} = 8 \pi GT_{\mu\nu} - \Lambda g_{\mu\nu}, \ee
with constant scalar profile $\phi=\phi_0=\text{const.}$ was derived in a covariant manner. 
The condition applies to any solution in GR with or without a matter component,
such as Schwarzschild, Kerr, de Sitter, and 
Friedmann-Lema\^itre-Robertson-Walker solutions, etc.
If we focus on the vacuum solution of the Lagrangian~\eqref{acC2}, the nontrivial conditions are 
\be \label{GRcon} F_0 + 2\Lambda F_2 =0 , \ee
at $\phi=\phi_0$ and the regularity of functions $F_i,A_I$ and their derivatives appearing in the Euler-Lagrange equations.

Here, we remark on the role of the degeneracy conditions.
In addition to 2 tensor degrees of freedom, 
the Lagrangian~\eqref{acC2} possesses a priori 2 scalar degrees of freedom, one of which is the Ostrogradsky ghost. 
To remove the Ostrogradsky ghost, one needs to impose a certain set of degeneracy conditions. 
Nevertheless, as we argued in Ref.~\cite{Motohashi:2018wdq} for GR solutions
in theories involving arbitrary higher-order derivatives,
we stress that the conditions for the existence of the particular black hole solutions 
are not directly related to the degeneracy conditions.
In \S\ref{sec3a} and \S\ref{sec3b}, we shall consider the Lagrangian~\eqref{acC2} without imposing the degeneracy conditions and derive exact solutions.
Therefore, strictly speaking, at this stage, we consider the most general Lagrangian~\eqref{acC2} quadratic in second-order derivatives of the scalar field, rather than the quadratic DHOST theories. 
In \S\ref{sec3deg}, we shall investigate all the classes of the quadratic DHOST theories and 
check the compatibility between the degeneracy conditions and the conditions to allow the exact solutions.

Let us proceed to the metric and scalar field ansatze. 
We consider the static and spherically symmetric spacetime 
\be \label{sss} ds^2 = - A(r)dt^2 + \f{dr^2}{B(r)} + 2C(r)dtdr+ D(r)r^2(d\theta^2+\cos^2\theta d\varphi^2) , \ee
with linearly time dependent profile of the scalar field
\be \label{phi-qt} \phi(t,r)=qt+\psi(r).\ee
From the fundamental theorem on gauge fixing at the action level~\cite{Motohashi:2016prk}, 
one can fix the gauge by imposing a gauge condition $D(r)=1$ at the action level without loss of independent Euler-Lagrange equation, 
while fixing the gauge by $C(r)=0$ at the action level does cause loss of independent Euler-Lagrange equation $\E_C=0$.
To avoid an inconsistent analysis of dynamics caused by the loss of independent Euler-Lagrange equation,
the gauge fixing $C(r)=0$ should be substituted after deriving the Euler-Lagrange equation.
It is a crucial difference from the standard case with the static scalar field in the static spherically symmetric spacetime, for which one can impose both gauge conditions $D(r)=1$ and $C(r)=0$ at the action level legitimately~\cite{Motohashi:2016prk}.
It is also worthwhile to remark that the argument is independent of the form of the action.
We will derive all the Euler-Lagrange equations $\E_A,\E_B,\E_C,\E_D,\E_\psi$ from the action, and then impose the gauge fixing condition $C(r)=0$ and $D(r)=1$.
While $\E_D=0$ and $\E_\psi=0$ can be recovered from other equations, $\E_D=0$ yields a simpler form and will be useful.
After the gauge fixing, the kinetic term reads
\be \label{kin}
X(r) = -\f{q^2}{A} + B \psi'^2,
\ee
where a prime denotes a derivative with respect to $r$.

From now on, we focus on the shift-symmetric subclass of the quadratic DHOST, where
\be F_i = F_i(X), \quad A_I = A_I(X)  \quad  (i=0,1,2;\,  I=1,\cdots, 5).  \ee
The shift symmetry simplifies the equations of motion dramatically.

Further, we impose an ansatz of constant canonical kinetic term
\be X=X_0 =\text{const.} \ee
With \eqref{kin}, this equation can be solved for $\psi'$ as
\be \label{psipsol} \psi'=\pm \sqrt{\f{q^2+X_0A}{AB}}, \ee
and hence higher derivatives of $\psi$ can be also written in terms of $q,X_0,A(r),B(r)$, and derivatives of $A(r),B(r)$.
As shown in \cite{Babichev:2013cya,Kobayashi:2014eva}, if $A,B$ are expanded as $A(r)=A_1 (r-r_h) + \cdots$ and $B(r)=B_1 (r-r_h) + \cdots$ at the vicinity of the (future/past) event horizon $r=r_h$, one can integrate \eqref{psipsol} and obtain
\be 
\psi = \pm \f{q}{\sqrt{A_1B_1}} \log \left| \f{r-r_h}{r_h} \right| + O (|r-r_h|) .
\ee
By choosing the plus sign of the two possible cases, and using the ingoing Eddington$-$Finkelstein coordinates $(v, r)$ defined by $dv = dt + dr/\sqrt{AB}$, one obtains $\phi\simeq qv$ at the vicinity of
the future event horizon.
Hence the scalar field is regular at the future event horizon.
On the other hand, if one chooses the minus branch of \eqref{psipsol},
one obtains the scalar field solution regular at the past event horizon.
In the rest, we shall focus on the black hole solutions and 
choose the plus sign of \eqref{psipsol}.

We shall write down the Euler-Lagrange equations by plugging the Schwarzschild or Schwarzschild-(anti-)de Sitter solutions to $A(r), B(r)$, and obtain sets of conditions for the coupling functions under which these solutions are allowed as exact solutions.

\section{Exact solutions}
\label{sec3}

In this section, we derive the conditions that guarantee the existence
of the Schwarzschild or Schwarzschild-(anti-)de Sitter solutions
in the most general Lagrangian~\eqref{acC2} quadratic in second-order derivatives of the scalar field
and provide general form of the coupling functions that satisfy the conditions.
We stress that at this stage we do not impose the degeneracy conditions and hence our analysis applies to all the class of the quadratic DHOST theories.

\subsection{Schwarzschild solution}
\label{sec3a}

First, let us focus on the Schwarzschild solution,
\be A=B= 1-\f{r_g}{r} =:f(r), \ee
where $r_g$ is constant, which corresponds to the mass of the Schwarzschild spacetime
and the event horizon $r_h=r_g$.
By integrating \eqref{psipsol}, the profile of the scalar field is given by
\be \label{psiqsol} 
\psi(r) = r \sqrt{Q} 
+ \f{(2 q^2 + X_0)r_g}{\sqrt{Q_0}} \log \kk{ \sqrt{r Q_0} \mk{ \sqrt{Q_0} + \sqrt{Q} } }
- 2 q r_g \,{\rm arctanh} \mk{ \f{q}{\sqrt{Q}} } 
+\psi_0, 
\ee
where
$\psi_0$ is an arbitrary integration constant, 
which has no physical importance in the shift symmetric theories, and 
\be Q(r) := q^2 + X_0 f(r), \quad 
Q_0 := q^2 + X_0. \ee
The Euler-Lagrange equations for $A,B,C,D,\psi$ are given by
\begin{align}
\E_A 
&= \f{X_0}{Q} Q_0 A_1 - \f{q^2}{Q} \E_B 
+  \f{q}{\sqrt{Q} f} \EC + \f{X_0}{2 Q} \ED ,\\
\E_B &= \f{1}{f}\mk{ \f{9 Q^2 + Q_0^2}{2 Q} - Q_0} (A_1 + A_2) 
- \f{1}{f}\mk{ Q_0A_1
+ \f{1}{2}\ED} \notag\\
&\quad + \f{Q}{2f^2} \biggl[ 2 r^2 F_{0X} 
+ \f{r (3 Q + Q_0)}{Q^{1/2}} F_{1X} 
+ \f{(3 Q + Q_0)^2}{2 Q} (A_{1X}+A_{2X}) 
- 2 Q_0 (2 A_{1X} + A_3 ) \biggr] , \\
\EC
&= \f{q}{\sqrt{Q}} \kk{ 2 Q_0 A_1 - \mk{ \f{9 Q^2 + Q_0^2}{2 Q} - Q_0 } (A_1 + A_2) + 2 f \E_B + \ED }, \\
\ED &= r^2 F_0 + \f{(9 Q - Q_0) (Q - Q_0)}{4 Q} (A_1 + A_2), \\
\E_\psi &= -\f{r (3 Q + Q_0)}{Q^{1/2}} F_{0X} 
-  4 Q_0 F_{1X} 
+ \f{ (Q-Q_0) [ 27 Q^3 - (11 q^2 + 2 X_0) Q^2 - (3 q^2 + X_0) Q_0 Q + 3 q^2 Q_0^2 ]}{4 r X_0 Q^{5/2}} (A_1 + A_2) \notag\\
&\quad + \f{ (9 Q - Q_0) (3 Q + Q_0) (Q - Q_0)}{4 r Q^{3/2}} ( A_{1X}+ A_{2X})
- \f{Q-Q_0}{r Q^{1/2}} Q_0 (2A_{1X}+A_3),
\end{align}
where 
$h_X := \pa h/\pa X$ for any function $h=h(X)$,
and $F_i$'s, $A_I$'s and their derivatives are evaluated at $X=X_0$.
As mentioned above, after deriving the Euler-Lagrange equations, we impose the gauge fixing conditions $C(r)=0$ and $D(r)=1$, which is the reason why $C,D$ and their derivatives do not appear in the above equations.
Here, we used the plus sign of the two cases of the solution \eqref{psipsol}.
The logic below is unchanged for the other case.

To make these equations vanish, the following set of conditions needs to be satisfied at $X=X_0$:
\be \label{con-Sch} F_0 = F_{0X} = F_{1X} = Q_0A_1 = A_1 + A_2 = A_{1X}+ A_{2X} = Q_0 (2A_{1X} + A_3) = 0 . \ee
Note that the conditions $A_1 + A_2=0$ and $A_{1X} + A_{2X}=0$ are satisfied by one of the degeneracy conditions for the Class I of the quadratic DHOST theories, i.e.\ $A_1(X)+A_2(X)=0$
which includes Horndeski and GLPV theories as a subclass.
There exist two possible branches, depending on whether $Q_0 =0$ or $Q_0 \ne 0$.

For $Q_0=0$, which we denote Case 1, the set of conditions reduces to 
\be \label{case-1} \text{Case 1}: \quad Q_0 = 0, \quad F_0 = F_{0X} = F_{1X} = A_1 + A_2 = A_{1X}+ A_{2X} = 0 . \ee

On the other hand, for $Q_0\ne 0$, which we denote Case 2, the set of conditions reduces to 
\be \label{case-2} \text{Case 2}: \quad Q_0\ne 0, \quad F_0 = F_{0X} = F_{1X} = A_1 = A_2 = A_{1X}+ A_{2X} = 2A_{1X} + A_3 = 0 . \ee
Note that $r_g$ does not appear in the conditions \eqref{case-1} and \eqref{case-2},
and the properties of the Schwarzschild metric is independent of the model parameters.
This is why the Schwarzschild solution in this type of models is called the {\it stealth} Schwarzschild solution.

Let us consider the case of the static scalar field $\phi=\psi(r)$ for which $q=0$,
for which the profile of the scalar field~\eqref{psiqsol} simplifies as
\be 
\psi(r) = \sqrt{X_0} \mk{r \sqrt{f} + r_g \log\kk{ \sqrt{r} \mk{ 1 + \sqrt{f} }} }+\psi_0, 
\ee
where 
$\psi_0$ is an arbitrary integration constant and $X_0>0$.
Assuming the regularity of $F_i$ and $A_I$ at $q= 0$, the Euler-Lagrange equations on the Schwarzschild metric ansatz reduce to 
\begin{align}
\E_A &= \f{1}{f} X_0 A_1 + \f{1}{2f} \ED ,\\
\E_B &= 
\f{X_0  (9 f^2 - 2 f + 1)}{2 f^2} (A_1 + A_2)
- \f{1}{f} X_0 A_1
- \f{1}{2f} \ED \\
&\quad + \f{1}{f}r^2 X_0 F_{0X} 
+ \f{r (3 f + 1)}{2f^{3/2}} X_0^{3/2} F_{1X} 
+ \f{(3 f + 1)^2}{4 f^2} X_0^2 (A_{1X}+A_{2X}) 
- \f{1}{f} X_0^2 (2 A_{1X} + A_3 )  ,\notag\\
\EC &= 0,\\
\ED &= r^2 F_0 + \f{(9 f - 1) (f - 1)}{4 f} X_0(A_1 + A_2),\\
\E_\psi &= 
- 4 X_0 F_{1X} 
- \f{r (1 + 3 f)}{ f^{1/2}} X_0^{1/2} F_{0X}
+ \f{(f - 1) (27 f^2 - 2 f - 1) }{4 r f^{3/2}} X_0^{1/2} (A_1 + A_2)
\notag\\
&\quad + \f{(9 f - 1) (3 f + 1) (f - 1) }{4 r f^{3/2}} X_0^{3/2} (A_{1X} + A_{2X}) 
- \f{f - 1}{r  f^{1/2}} X_0^{3/2} (2 A_{1X} + A_3) ,
\end{align}
which are consistent with the background equations for the static spherically symmetric configuration in the Horndeski theory~\cite{Kobayashi:2012kh,Kobayashi:2014wsa}.

For $X_0\ne 0$, the condition coincides with \eqref{case-2} with $Q_0\to X_0$.
On the other hand, for the constant profile $\phi=\phi_0=\text{const.}$ for which $X_0=0$, the condition is simply given by
\be \label{case-1c} \text{Case 1-c}: \quad  \phi=\phi_0, \quad F_0 = 0 , \ee
at $X=0$.
This is consistent with the condition~\eqref{GRcon} with $\Lambda=0$.

Besides the exact Schwarzschild solutions,
there also exists the Schwarzschild-type solution
with a deficit solid angle
$A(r)=f(r)=1-r_g/r$
and 
$B(r)=\alpha f(r)$ ($\alpha \neq 1$),
for the models satisfying
\be \label{Sch_type}
\alpha=\frac{F_2}{F_2-X_0 A_1}, \quad
2A_1F_{2X}=-F_2(2A_{1X}+A_3), \quad
q=F_{0}=F_{0X}=F_{1X}=0, \quad
A_{2}=-A_{1}, \quad
A_{2X}=-A_{1X}.
\ee
Unless $A_1=0$, $\alpha\neq 1$, and hence, there is a solid deficit angle.
Note that for $A_1=0$ the condition \eqref{Sch_type} reduces to Case 2 \eqref{case-2} with $q=0$.
In this case the scalar field profile is given by
\be \psi(r) = \sqrt{\f{X_0}{\alpha}} \mk{r \sqrt{f} + r_g \log\kk{ \sqrt{r} \mk{ 1 + \sqrt{f} }} }+\psi_0, \ee
$\psi_0$ is an arbitrary integration constant and $X_0>0$.
It is interesting to note that 
in the quadratic DHOST theories with $c_{t}=c$ where $A_1=0$ (see \S\ref{sec3d}),
$\alpha=1$, and hence, no solid deficit angle exists.

\subsection{Schwarzschild-(anti-)de Sitter solution}
\label{sec3b}

Next let us focus on Schwarzschild-(anti-)de Sitter solutions
\be A=B= 1-\f{r_g}{r}-\f{\Lambda}{3}r^2 =:f_\Lambda(r), \ee
where $r_g$ and $\Lambda$ are constants ($r_h\neq r_g$).
The case $\Lambda>0$ ($\Lambda<0$) corresponds 
to the case of the asymptotically de Sitter (anti-de Sitter) solution.
In this case, the Euler-Lagrange equations for $A,B,C,D, \psi$ are given by
\begin{align}
\E_{A,\Lambda} 
&= \f{X_0}{Q_\Lambda} Q_0 A_1 - \f{q^2}{Q_\Lambda} \E_{B,\Lambda} 
+ \f{q}{\sqrt{Q_\Lambda} f_\Lambda} \ELC + \f{X_0}{2 Q_\Lambda} \ELD ,\\
\E_{B,\Lambda} &= 
\f{1}{f_\Lambda} \mk{ \f{ 9 Q_\Lambda^2 + Q_0^2 + 2 \lambda (3 Q_\Lambda - Q_0) + \lambda^2}{2 Q_\Lambda} - Q_0 } (A_1 + A_2) 
- \f{1}{f_\Lambda} \mk{ Q_0 A_1 
+ \f{1}{2} \ELD } \notag\\
&\quad + \f{Q_\Lambda}{2f_\Lambda^2} \biggl[ r^2 [2 F_{0X} + \Lambda (8 F_{2X} - 2A_1 + 4 X_0 A_{1X} + 3 X_0 A_3 )]  
+ \f{r (3 Q_\Lambda + Q_0 - \lambda)}{\sqrt{Q_\Lambda}} F_{1X}  \notag\\
&\quad
+ \f{(3 Q_\Lambda + Q_0)^2 - \lambda (6 Q_\Lambda + 2 Q_0 - \lambda)}{2 Q_\Lambda} (A_{1X} + A_{2X}) 
- 2 Q_0  (2 A_{1X} + A_3 ) \biggr] , \\
\ELC
&= \f{q}{\sqrt{Q_\Lambda}} \kk{ 2 Q_0 A_1 - \mk{ \f{ 9 Q_\Lambda^2 + Q_0^2 + 2 \lambda (3 Q_\Lambda - Q_0) + \lambda^2}{2 Q_\Lambda} - Q_0 } (A_1 + A_2) + 2 f_\Lambda \E_{B,\Lambda} + \ELD }, \\
\ELD &= 
r^2 [F_0 + 2 \Lambda (F_2 - X_0 A_1)]
+ \f{1}{4 Q_\Lambda} [(9 Q_\Lambda - Q_0) (Q_\Lambda - Q_0 + 2 \lambda) + \lambda^2] (A_1 + A_2)  ,\\
\E_{\psi,\Lambda} &= 
-\f{r (3 Q_\Lambda + Q_0 - \lambda)}{2 Q_\Lambda^{1/2} } [2 F_{0X} + \Lambda (8 F_{2X} - 2A_1 + 4 X_0 A_{1X} + 3 X_0 A_3 )]
- 2 (2 Q_0 - 3 \lambda) F_{1X} \notag\\
&\quad + \f{1}{4 r X_0 Q_\Lambda^{5/2}} \biggl[ 
(Q_\Lambda - Q_0) [ 27 Q_\Lambda^3 - (11 q^2 + 2 X_0) Q_\Lambda^2 - (3 q^2 + X_0) Q_0 Q_\Lambda + 3 q^2 Q_0^2 ] \notag\\
&\quad + \lambda \Bigl\{ 9 Q_\Lambda^3 + (23 q^2 + 2 X_0) Q_\Lambda^2 - Q_0 (25 q^2 + 3 X_0) Q_\Lambda +  9 q^2 Q_0^2 \notag\\
&\quad + [-3 Q_\Lambda^2 + (20 q^2 + 3 X_0) Q_\Lambda - 9 Q_0 q^2] \lambda 
+ (3 q^2 - Q_\Lambda) \lambda^2 \Bigr\} \biggr] (A_1 + A_2) \notag\\
&\quad + \f{1}{4 r Q_\Lambda^{3/2}  } \biggl[(9 Q_\Lambda - Q_0) (3 Q_\Lambda + Q_0) (Q_\Lambda - Q_0) 
+ \lambda \Bigl\{ (9 Q_\Lambda - Q_0) (5 Q_\Lambda + 3 Q_0) - 3 (5 Q_\Lambda - Q_0) \lambda - \lambda^2 \Bigr\}\biggr] (A_{1X} + A_{2X}) \notag\\
&\quad - \f{Q_\Lambda - Q_0 + \lambda}{r Q_\Lambda^{1/2}} Q_0 (2 A_{1X} + A_3) ,
\end{align}
where
\be 
Q_\Lambda(r) := q^2 + X_0 f_\Lambda(r), \quad 
Q_0 := q^2 + X_0, \quad 
\lambda(r) := X_0 \Lambda r^2 .
\ee
Again, $F_i$'s, $A_I$'s and their derivatives
are evaluated at $X=X_0$, and the signs correspond to the two cases of the solution \eqref{psipsol}. 
These equations reduce to the case for the Schwarzschild solution for $\Lambda\to 0$.

To make these equations vanish, the following set of conditions needs to be satisfied at $X=X_0$:
\begin{align} \label{con-SdS} 
&F_0 + 2 \Lambda (F_2 - X_0 A_1) 
= 2 F_{0X} + \Lambda (8 F_{2X} - 2A_1 + 4 X_0 A_{1X} + 3 X_0 A_3 )  \notag\\
&= F_{1X} = Q_0 A_1 = A_1 + A_2 = A_{1X}+ A_{2X} = Q_0 (2 A_{1X} + A_3) =0, 
\end{align}
which is a generalization of the condition~\eqref{con-Sch} for nonzero $\Lambda$, and applies for general $Q_0$ and $\Lambda$.
Again, the conditions $A_1 + A_2=0$ and $A_{1X}+A_{2X}=0$ are satisfied by the degeneracy condition for Class I of the quadratic DHOST theories.
Depending on whether $Q_0 =0$ or $Q_0 \ne 0$, there are two branches, 
which we denote Case 1-$\Lambda$ and Case 2-$\Lambda$,
\begin{align} \label{case-1L} 
\text{Case 1-}\Lambda: \quad Q_0 =0, \quad &F_0 + 2 \Lambda (F_2 + q^2 A_1) 
= 2 F_{0X} + \Lambda [8 F_{2X} - 2A_1 - q^2 (4  A_{1X} + 3 A_3 ) ]  \notag\\
&= F_{1X} = A_1 + A_2 = A_{1X}+ A_{2X} = 0,  \\
\label{case-2L} 
\text{Case 2-}\Lambda: \quad  Q_0\ne 0, \quad  &F_0 + 2 \Lambda F_2 
= F_{0X} + \Lambda (4 F_{2X} - X_0 A_{1X} ) \notag\\
&= F_{1X} = A_1 = A_2 = A_{1X}+ A_{2X} = 2 A_{1X} + A_3 =0.
\end{align}
Note that the limit $\Lambda\to 0$ of these branches precisely recover the two branches: 
Case 1~\eqref{case-1} and 
Case 2~\eqref{case-2} for the Schwarzschild solution. 
Note also that since $\Lambda$ appears in the conditions \eqref{case-1L} and \eqref{case-2L},
the Schwarzschild-(anti-)de Sitter solution is not of the stealth type,
in contrast to the case of the Schwarzschild solution in \S\ref{sec3a}.

Let us consider the static case $\phi=\psi(r)$ for which $q=0$.
Assuming the regularity of $F_i$, $A_I$ and their derivatives
at $q= 0$, the Euler-Lagrange equations on the Schwarzschild-de Sitter metric ansatz reduce to 
\begin{align}
\E_{A,\Lambda} &= \f{1}{f_\Lambda} X_0 A_1 + \f{1}{2f_\Lambda} \ELD ,\\
\E_{B,\Lambda} &= 
\f{9 f_\Lambda^2 - 
 2 (1 - 3 \Lambda r^2 ) f_\Lambda + (1 - \Lambda r^2 )^2}{2 f_\Lambda^2} X_0(A_1 + A_2)
- \f{1}{f_\Lambda} X_0 A_1
- \f{1}{2f_\Lambda} \ELD \notag\\
&\quad + \f{r^2 }{2f_\Lambda} X_0 [2 F_{0X} + \Lambda (8 F_{2X} - 2A_1 + 4 X_0 A_{1X} + 3 X_0 A_3 )]  
+ \f{r (3 f_\Lambda + 1 - \Lambda r^2)}{2f_\Lambda^{3/2}} X_0^{3/2} F_{1X}  \notag\\
&\quad
+ \f{(3 f_\Lambda + 1 - \Lambda r^2)^2}{4 f_\Lambda^2} X_0^2(A_{1X} + A_{2X}) 
- \f{1}{f_\Lambda} X_0^2  (2 A_{1X} + A_3 ) ,\\
\ELC
&= 0,\\
\ELD &= r^2 [F_0 + 2 \Lambda (F_2 - X_0 A_1)]
+ \f{ (9 f_\Lambda-1) (f_\Lambda-1 + 2 \Lambda r^2) + \Lambda^2r^4 }{4 f_\Lambda} X_0(A_1 + A_2),\\
\E_{\psi,\Lambda} &= 
-\f{r (3 f_\Lambda + 1 - \Lambda r^2)}{2 f_\Lambda^{1/2} } X_0^{1/2} [2 F_{0X} + \Lambda (8 F_{2X} - 2A_1 + 4 X_0 A_{1X} + 3 X_0 A_3 )]
- 2 (2 - 3 \Lambda r^2) X_0 F_{1X} \notag\\
&\quad + \f{1}{4 r f_\Lambda^{3/2}} \biggl[ 
27 f_\Lambda^3 - (29 - 9 \Lambda r^2) f_\Lambda^2 + (1 - \Lambda r^2) (1 + 3 \Lambda r^2) f_\Lambda + (1 - \Lambda r^2)^3 \biggr] X_0^{1/2} (A_1 + A_2) \notag\\
&\quad + \f{1}{4 r f_\Lambda^{3/2}  } \biggl[ 27 f_\Lambda^3 - 
 3 (7 - 15 \Lambda r^2) f_\Lambda^2 - (1 - \Lambda r^2) (7 - 15 \Lambda r^2) f_\Lambda + (1 - \Lambda r^2)^3 \biggr] X_0^{3/2} (A_{1X} + A_{2X}) \notag\\
&\quad - \f{f_\Lambda - 1 + \Lambda r^2}{r f_\Lambda^{1/2}} X_0^{3/2} (2 A_{1X} + A_3) ,
\end{align}
which are consistent with the background equations for the static spherically symmetric configuration in the Horndeski theory~\cite{Kobayashi:2012kh,Kobayashi:2014wsa}.

For $X_0\ne 0$, the condition coincides with \eqref{case-2L} with $Q_0\to X_0$.
On the other hand, for the constant scalar profile $\phi=\text{const.}$ for which $X_0=0$, the condition is simply given by
\be \label{case-1Lc} \text{Case 1-}\Lambda\text{-c}: \quad \phi=\phi_0, \quad F_0 + 2 \Lambda F_2  = 0 , \ee
at $X=0$.

\subsection{Degeneracy conditions}
\label{sec3deg}

So far we have not imposed the degeneracy conditions, which are responsible for eliminating the Ostrogradsky ghost.
Now let us consider the degeneracy conditions and study their compatibility with the conditions to allow the exact solutions.
The quadratic DHOST theories are classified into several classes depending on how they satisfy three degeneracy conditions~\cite{Langlois:2015cwa}.

Class I is defined by $A_1+A_2=0$, which is compatible with the condition for all the cases considered above.
Further, the remaining degeneracy conditions can be satisfied in two ways:
subclass Ia is defined by $F_2\ne XA_1$, and $A_4, A_5$ are written down in terms of $A_2, A_3$,
and subclass Ib is defined by $F_2=XA_1$ and $A_3=2(F_2-2XF_{2X})/X^2$.
These conditions are also compatible with all the conditions obtained above.

On the other hand, Class II is defined by $F_2\ne 0$ and $A_1+A_2\ne 0$, which are 
compatible with the conditions for Cases 1-c, 1-$\Lambda$-c, but
incompatible with the conditions for Cases 1, 1-$\Lambda$, 2, 2-$\Lambda$.
Therefore, so long as we focus on the ansatz of the linearly time dependent profile~\eqref{phi-qt} of the scalar field,
for the shift-symmetric subclass of Class II, 
the Schwarzschild or Schwarzschild-(anti-)de Sitter solution
is allowed only for a constant scalar field,
and not for the linearly time dependent profile with $q\ne 0$ and/or nontrivial radial profile $\psi'(r)\neq 0$.

Finally, Class III is defined by $F_2=0$, which is compatible with the condition for all the cases considered above.
Subclass IIIa requires $A_1+3A_2\ne 0$, which is compatible with Cases 1, 1-$\Lambda$, 1-$\Lambda$-c, but incompatible with Cases 2, 2-$\Lambda$ as they require $A_1(X_0)=A_2(X_0)=0$.
Subclasses IIIb and IIIc, respectively, require $A_1+3A_2=0$ and $A_1=0$, which are compatible with all the cases.

\subsection{Models}
\label{sec3c}

Let us provide general models that satisfy the conditions.
Since we assume the regularity at $X=X_0$, general forms of 
$F_i$'s and $A_I$'s can be given in terms of Taylor series,
\be \label{taylorexp} F_i(X) = \sum_{n=0}^\infty f_{in} (X-X_0)^n , \quad 
A_I(X) = \sum_{n=0}^\infty a_{In} (X-X_0)^n, \quad  (i=0,1,2;\,  I=1,\cdots, 5), \ee
From the conditions, we obtain equations between the coefficients $f_{in},a_{In}$.

Specifically, for the Schwarzschild solution,
\begin{align} 
\label{case-1-co} \text{Case 1}: \quad Q_0=0, \quad &f_{00} = f_{01} = f_{11} = 0, \quad a_{20} = - a_{10} ,\quad a_{21} = - a_{11} ,\\
\label{case-2-co} \text{Case 2}: \quad Q_0\ne 0, \quad &f_{00} = f_{01} = f_{11} = a_{10} = a_{20} = 0, \quad a_{21} = -a_{11}, \quad a_{30} = -2a_{11},\\
\label{case-1c-co} \text{Case 1-c}: \quad \phi=\phi_0, \quad &f_{00} = 0 ,
\end{align} 
and for the Schwarzschild-de Sitter solution,
\begin{align} 
\label{case-1L-co} 
\text{Case 1-}\Lambda: \quad Q_0 = 0, \quad 
&f_{00} = - 2 \Lambda (f_{20} + q^2 a_{10}), \quad  
f_{01} = - \f{\Lambda}{2} [8 f_{21} - 2a_{10} -  q^2(4 a_{11} + 3 a_{30}) ] , \notag\\
&f_{11} = 0,\quad 
a_{20} = - a_{10},\quad 
a_{21} = - a_{11}, \\
\label{case-2L-co} 
\text{Case 2-}\Lambda: \quad  Q_0\ne 0, \quad  
&f_{00} = - 2 \Lambda f_{20},\quad 
f_{01} = - \Lambda (4 f_{21} - X_0 a_{11} ), \notag\\ 
&f_{11} = a_{10} = a_{20} =0,\quad a_{21}= -a_{11}, \quad a_{30} = -2 a_{11}, \\
\label{case-1Lc-co} 
\text{Case 1-}\Lambda\text{-c}: \quad \phi=\phi_0, \quad &f_{00} = - 2 \Lambda f_{20} , 
\end{align}
For each case, other coefficients that do not appear in the above conditions are arbitrary.

\subsection{Scalar-tensor theories with $c_t=c$ and without decay of gravitons}
\label{sec3d}

For the quadratic DHOST theories, the propagation speed $c_t$ of gravitational waves coincides with the speed of light $c$ by imposing the additional condition~\cite{Langlois:2017dyl},
\be \label{ctc} A_1=A_2=0, \quad A_4 = \f{1}{8F_2}[48F_{2X}^2 -8(F_2-XF_{2X}) A_3 - X^2A_3^2], \quad A_5 = \f{1}{2F_2}(4F_{2X}+XA_3)A_3 .  \ee
This subclass is interesting from the point of view of the stringent constraint on the deviation $c_t-c$ on cosmological scales down to the accuracy of order $10^{-15}$ imposed by the LIGO/Virgo and optical observations of a binary neutron star merger \cite{TheLIGOScientific:2017qsa,GBM:2017lvd,Monitor:2017mdv}.
For the quadratic DHOST subclass in which $c_t=c$, the condition~\eqref{ctc} needs to be satisfied in addition to the conditions for each cases.\footnote{\label{note1}Recently, black hole solutions with linearly time dependent scalar with $X^2+q^2=0$ for the subclass of the quadratic DHOST theories 
with the condition $A_1(\phi,X)=A_2(\phi,X)=0$
was studied in \cite{BenAchour:2018dap}.
Here, note that in \cite{BenAchour:2018dap} the notation is $X := - \phi_{\mu}\phi^{\mu}/M^2$ and the Schwarzschild-(anti-)de Sitter metric is defined as $-g_{tt}=1/g_{rr}=1-2m/r-\Lambda r^2$.
They obtain two sets of conditions, Eqs.~(28), (29) for the Schwarzschild solution, and Eqs.~(22) -- (24) for the Schwarzschild-(anti-)de Sitter solution, which would amount to the Case 1~\eqref{case-1} and Case 1-$\Lambda$~\eqref{case-1L}, respectively. 
However, their results are different from our conditions:
(i) $F_{0X}=0$ in our \eqref{case-1} is absent in Eqs.~(28), (29) of \cite{BenAchour:2018dap},
and 
(ii) the condition (24) in \cite{BenAchour:2018dap} is absent in our \eqref{case-1L}.}
For the subclass with $c_t=c$, 
by plugging \eqref{taylorexp} into \eqref{ctc} and evaluating each order, we find equations between $f_{2n},a_{3n},a_{4n},a_{5n}$.
For instance, from the coefficient for $(X-X_0)^0$ order, we obtain 
\begin{align}
\label{a4}
a_{40}
&=\f{1}{8f_{20}}[48 f_{21}^2- 8(f_{20}-8X_0f_{21}) a_{30}-X_0^2 a_{30}^2], 
\\
\label{a5}
a_{50}
&=\f{1}{2f_{20}}(4f_{21}+X_0 a_{30})a_{30}.
\end{align}
It is also straightforward to obtain similar equations for $a_{4n},a_{5n}$ for $n\geq 1$. 
Since $A_4$ and $A_5$ do not appear in the conditions 
for the existence of the Schwarzschild and the Schwarzschild-(anti-)de Sitter solutions,
the relations \eqref{a4} and \eqref{a5} (and those for $a_{4n},a_{5n}$ for $n\geq 1$)
do not affect \eqref{case-1-co} and \eqref{case-1Lc-co}.

Furthermore, it was claimed in \cite{Creminelli:2018xsv} that the additional condition 
\be \label{dec} A_3=0, \ee 
should be satisfied to prevent a rapid decay of gravitational waves into dark energy fluctuations.
After imposing the additional condition~\eqref{dec}, free coupling functions are $F_0,F_1,F_2$.
Note that with the additional condition~\eqref{dec} rules out GLPV theories and Horndeski or ``pure'' DHOST theories that do not include GLPV as a subclass survive (see \S\ref{sec4a}).
The action then takes the following form:
\be \label{qdrem} S = \int d^4x \sqrt{-g} \kk{ F_0 + F_1 \Box\phi + F_2 R + \f{6F_{2X}^2}{F_2}\phi^\mu \phi_{\mu\nu} \phi^{\nu\lambda} \phi_\lambda }. \ee

A caveat on the constraints~\eqref{ctc} and \eqref{dec} is that the energy scales observed by LIGO/Virgo is close to the cutoff scale of the effective field theory of dark energy, and future observations of lower frequency gravitational waves is necessary for more conclusive constraints on the models of dark energy~\cite{deRham:2018red}.
Moreover, in general the constraint does not necessarily apply to the models for the early Universe or the vicinity of black holes, whose energy scales are higher than the LIGO/Virgo constraints.

\section{Special cases}
\label{sec4}

In \S\ref{sec3a} and \S\ref{sec3b}, we derived the conditions in the shift-symmetric quadratic DHOST theories.
In this section we write down conditions for several subclasses, namely, the Horndeski and GLPV theories, as well as the subclass in which $c_t=c$ by imposing \eqref{ctc}.

\subsection{Horndeski and GLPV theories}
\label{sec4a}

The quadratic DHOST includes a subclass of GLPV theory up to quartic order interaction (i.e.\ without $G_5, F_5$), which is characterized by 
\be \label{GLPV} 
F_0= G_2, \quad
F_1=G_3, \quad 
F_2=G_4, \quad 
A_1 = - A_2 = 2 G_{4X} + Xf_4, \quad 
A_3= -A_4 = 2f_4, \quad A_5=0. \ee
As a special case, the Horndeski theory up to quartic order interaction (i.e.\ without $G_5$) can be obtained by imposing \eqref{GLPV} and an additional condition 
\be \label{Horn} f_4=0. \ee 
Note that these conditions are imposed for any $(\phi,X)$.

For the GLPV subclass~\eqref{GLPV} with shift symmetry, the conditions at $X=X_0$ for 
Case 1~\eqref{case-1}, 
Case 2~\eqref{case-2}, and 
Case 1-c~\eqref{case-1c} 
for the Schwarzschild solution, respectively, reduce to 
\begin{align} 
\label{case-1-G} 
\text{Case 1}: \quad Q_0=0, \quad &G_2 = G_{2X} = G_{3X} = 0 , \\
\label{case-2-G} 
\text{Case 2}: \quad Q_0\ne 0, \quad &G_2 = G_{2X} = G_{3X} = 2 G_{4X} + X_0 f_4 = 2 G_{4XX} + 2 f_4 + X_0f_{4X} = 0 ,\\
\label{case-1c-G} 
\text{Case 1-c}: \quad \phi=\phi_0, \quad &G_2 = 0 . 
\end{align}
Case 2 includes the stealth Schwarzschild solution considered in \cite{Babichev:2017guv} [see Eq.~(5.3)]. 
Likewise, the conditions at $X=X_0$ for 
Case 1-$\Lambda$~\eqref{case-1L},
Case 2-$\Lambda$~\eqref{case-2L}, and 
Case 1-$\Lambda$-c \eqref{case-1Lc} 
for the Schwarzschild-(anti-)de Sitter solution respectively reduce to
\begin{align} 
\label{case-1L-G} 
\text{Case 1-}\Lambda: \quad 
&Q_0 = 0, \notag\\
&G_2 + 2 \Lambda (G_4 + 2 q^2 G_{4X} - X_0^2f_4)
= G_{2X} + 2 \Lambda [ G_{4X} - 2 q^2( G_{4XX} + f_4) + q^4 f_{4X} ] 
= G_{3X} = 0 ,\\
\label{case-2L-G} 
\text{Case 2-}\Lambda: \quad 
&Q_0\ne 0,  \notag\\
&G_2 + 2 \Lambda G_4 
= G_{2X} + \Lambda (4 G_{4X} + X_0 f_4 )
= G_{3X} 
= 2 G_{4X} + X_0f_4 
= 2 G_{4XX} + 2f_4 + X_0 f_{4X} =0, \\
\label{case-1Lc-G} 
\text{Case 1-}\Lambda\text{-c}: \quad
&\phi=\phi_0, \quad G_2 + 2 \Lambda G_4 = 0 . 
\end{align}

\subsection{Scalar-tensor theories with $c_t=c$ and without decay of gravitons}
\label{sec4b}

For the subclass that satisfies the condition~\eqref{ctc} and the shift symmetry, the conditions for the existence of the Schwarzschild solution reads
\begin{align} 
\label{case-1-ctc} \text{Case 1}: \quad Q_0 =0, \quad  &F_0 = F_{0X} = F_{1X} =  0 ,\\
\label{case-2-ctc} \text{Case 2}: \quad Q_0\ne 0, \quad &F_0 = F_{0X} = F_{1X} = A_3 = 0 ,\\
\label{case-1c-ctc} \text{Case 1-c}: \quad  \phi=\phi_0, \quad &F_0 = 0 , 
\end{align} 
and for the Schwarzschild-de Sitter solution
\begin{align} 
\label{case-1L-ctc} 
\text{Case 1-}\Lambda: \quad Q_0 =0,\quad 
&F_0 + 2 \Lambda F_2 
= 2 F_{0X} + \Lambda (8 F_{2X} - 3q^2 A_3 ) 
= F_{1X} = 0, \\
\label{case-2L-ctc} 
\text{Case 2-}\Lambda: \quad  Q_0\ne 0, \quad  
&F_0 + 2 \Lambda F_2 
= F_{0X} + 4 \Lambda F_{2X} 
= F_{1X} = A_3 =0, \\
\label{case-1Lc-ctc}
\text{Case 1-}\Lambda\text{-c}: \quad \phi=\phi_0,\quad 
&F_0 + 2 \Lambda F_2  = 0 . 
\end{align}
It is interesting to note that for $A_1=0$
the Schwarzschild-type solution~\eqref{Sch_type} 
coincides with the exact Schwarzschild solution since $\alpha=1$,
and hence, no deficit solid angle exists for the models with $c_t=c$.
Furthermore, for models without graviton decay, 
the additional condition~\eqref{dec} needs to be imposed, which is already satisfied for Case 2~\eqref{case-2-ctc} and Case 2-$\Lambda$~\eqref{case-2L-ctc} for $c_t=c$.

For GLPV theory~\eqref{GLPV}, the additional condition~\eqref{ctc} for $c_t=c$ reduces to \cite{Creminelli:2017sry}
\be \label{ctc-G} 2 G_{4X} + Xf_4=0, \ee
for any $(\phi,X)$.
Imposing the shift symmetry, the conditions at $X=X_0$ for the existence of the Schwarzschild solution reads
\begin{align} 
\label{case-1-G-ctc} \text{Case 1}: \quad Q_0=0,\quad &G_2 = G_{2X} = G_{3X} = 0 ,\\
\label{case-2-G-ctc} \text{Case 2}: \quad Q_0\ne 0, \quad &G_2 = G_{2X} = G_{3X} = G_{4X} = f_4 = 0 ,\\
\label{case-1c-G-ctc} \text{Case 1-c}: \quad \phi=\phi_0, \quad  &G_2 = 0 , 
\end{align} 
and for the Schwarzschild-de Sitter solution
\begin{align} 
\label{case-1L-G-ctc} 
\text{Case 1-}\Lambda: \quad 
Q_0 = 0, \quad
&G_2 + 2 \Lambda G_4 
= G_{2X} - 2 \Lambda G_{4X}  
= G_{3X} = 0, \\
\label{case-2L-G-ctc} 
\text{Case 2-}\Lambda: \quad  
Q_0\ne 0, \quad 
&G_2 + 2 \Lambda G_4 
= G_{2X} 
= G_{3X} 
= G_{4X} 
= f_4 =0, \\
\label{case-1Lc-G-ctc} 
\text{Case 1-}\Lambda\text{-c}: \quad \phi=\phi_0,\quad 
&G_2 + 2 \Lambda G_4 = 0 ,
\end{align}
among which \eqref{case-2-G-ctc} and \eqref{case-2L-G-ctc} satisfy the Horndeski condition~\eqref{Horn}.
Also, as mentioned above, the additional condition~\eqref{dec} to prevent a rapid graviton decay means $f_4=0$, which makes the GLPV theory to Horndeski one.

Note that, for Horndeski theory, from \eqref{Horn} the condition \eqref{ctc-G} implies $G_4=G_4(\phi)$, and hence, the analysis in \S\ref{sec3} for the shift-symmetric theories applies only if $G_4=\text{const}$.

\section{Applications}
\label{sec5}

In \S\ref{sec3c}, we provided general models that satisfy each condition.
In this section, as another application of the conditions, we provide specific simple models that allow novel exact black hole solutions, which include known solutions and models in the literature as a special case.

\subsection{Constant scalar solution}
\label{sec5a}

The condition for any GR solution with constant scalar profile for a wide range of higher-derivative scalar-tensor theories was derived in \cite{Motohashi:2018wdq}.  
Cases 1-c and 1-$\Lambda$-c are precisely special cases of the condition derived there, i.e.\ \eqref{GRcon}.  
The general form of models that satisfy the condition is given in \S\ref{sec3c} in terms of Taylor series; i.e.\
\be \label{case-1Lc-ex1}
F_0 = - \Lambda \Mpl^2 + M^4 a(X), \quad
F_2 = \f{\Mpl^2}{2} + M^2 b(X), 
\ee
where 
$a(X) = \sum_{n=1}^\infty f_{0n} X^n, 
b(X) = \sum_{n=1}^\infty f_{2n} X^n$, 
and hence, $a(0)=b(0)=0$. 
Other coupling functions $F_i$ and $A_I$ are free functions so long as they are regular at $X=0$ together with their derivatives appearing in the Euler-Lagrange equations. 
For Case 1, $F_2$ can be also a free regular function. 
For Case 1-$\Lambda$, one may note that the condition $a(0)=b(0)=0$ is sufficient but not necessary.
Since the condition~\eqref{case-1Lc} for Case 1-$\Lambda$ requires $M^2 a(0)+2\Lambda b(0)=0$, $a(0)$ and $b(0)$ do not need to vanish.
A simple example of such case is derivatives appearing in the Euler-Lagrange equations, 
\be \label{case-1Lc-ex2}
F_0 = - \Lambda [ \Mpl^2 + M^2 h(X) ], \quad
F_2 = \f{\Mpl^2}{2} + \f{M^2}{2} h(X), 
\ee
with $h(0)\ne 0$.
In this model, even though $F_0(0)$ and $F_2(0)$ have nonvanishing $h(0)$, they cancel in the condition~\eqref{case-1Lc}.
Such a cancellation is only possible for Case 1-$\Lambda$-c with nonzero cosmological constant $\Lambda\ne 0$.

\subsection{Stealth Schwarzschild solution}
\label{sec5b}

The stealth Schwarzschild solutions in shift-symmetric Horndeski theory are obtained in \cite{Babichev:2013cya,Kobayashi:2014eva}.
The theory with 
$G_4=\Mpl^2/2 - \beta X/2$ with $\Mpl := (8\pi G)^{-1/2}$ being the reduced Planck mass
and $\beta$ being constant with mass dimension $-2$
is considered in \cite{Babichev:2013cya}, and the theory with 
$G_4=G_4(X)$ is considered in \cite{Kobayashi:2014eva}. 
These classes fall into Cases 1 and 2 for the shift-symmetric Horndeski, which are defined by the conditions~\eqref{case-1-G} and \eqref{case-2-G} after plugging \eqref{Horn}.

The Schwarzschild-type solution with a deficit solid angle
[Eq.~(5.2) in Ref.~\cite{Babichev:2017guv}],
$A(r)=1-r_g/r$ and $B(r)=\alpha A(r)$, 
is recovered
from the GLPV limit of the condition \eqref{Sch_type}
with
\be
\alpha = \frac{G_4}{G_4-X_0 (2G_{4X} + X_0 f_4)}.
\ee
In the shift-symmetric subclass with $c_{t}=c$,
for which Eq.~\eqref{ctc-G} holds, $\alpha=1$ and no solid deficit angle exists.

On the other hand, asymptotically flat black hole solutions
obtained in Refs.~\cite{Sotiriou:2013qea,Sotiriou:2014pfa,Babichev:2017guv}
in the shift-symmetric Horndeski and GLPV theories are not included in our analysis,
since in these solutions $X$ is not constant.
Also, the two stealth Schwarzschild solutions with $\phi=\phi(r)$ obtained in \cite{Minamitsuji:2018vuw} do not fall into our analysis since they exist in the shift-symmetry breaking subclass of the Horndeski theory and $X$ is not constant as well.
In this case, the scalar profile has only radial coordinate dependence $\phi=\phi(r)$.

In the case of the quadratic DHOST,
as mentioned in \S\ref{sec3a}
all the Schwarzschild solutions satisfying \eqref{case-1} or \eqref{case-2}
are of the stealth type and a broad class of the models with 
the regular functions $F_i$'s and $A_I$'s possesses these solutions.
The general models were given in \S\ref{sec3c}.
As an example, a simple model in the DHOST subclass with $c_t=c$
possessing the stealth Schwarzschild solution is given by 
\be F_0=M^4 a(X), \quad 
F_2= \frac{\Mpl^2}{2} + M^2 b(X), \quad
A_3= \frac{c(X)}{M^6},
\ee
with 
\be
A_1=A_2=0, \quad
A_4 = -A_3  
- \frac{[ Xc(X)+4M^8b'(X) ] [ Xc(X) -12M^8 b'(X) ]}
       {4M^{12} [ \Mpl^2+2M^2b(X) ]}, \quad
A_5= \frac{c(X)[ X c(X)+4M^8b'(X) ]}
         {M^{12} [ \Mpl^2+2M^2 b(X) ]},
\ee
where $a(X), b(X), c(X)$ are dimensionless regular functions of $X$
and $M$ is a constant with mass dimension 1.
The conditions for Case 1~\eqref{case-1} and 
Case 2~\eqref{case-2} are satisfied by imposing
$a(-q^2)=a'(-q^2)=F_{1X}(-q^2)=0$
and
$a(X_0)=a'(X_0)=F_{1X}(X_0)=c(X_0)=0$,
respectively.
Note that $b(X)$ remains as free function.
The condition~\eqref{dec} to prevent a rapid graviton decay is satisfied by imposing $c(X)=0$.
Note also that this model automatically satisfies the degeneracy condition $A_1+A_2=0$ for Class I.

\subsection{Schwarzschild-(anti-)de Sitter solution}
\label{sec5c}

The Schwarzschild-(anti-)de Sitter solutions in shift-symmetric Horndeski theory are obtained in \cite{Babichev:2013cya,Kobayashi:2014eva}.
It is shown in \cite{Babichev:2013cya} that the model defined by 
\be \label{b13}
G_2= -\Mpl^2 \Lambda_{\rm b}-\eta X, \quad  
G_4=\frac{\Mpl^2}{2} - \frac{\beta X}{2}, 
\ee
with $\Lambda_{\rm b}, \eta, \beta$ being constants 
with mass dimension $2,0,-2$ respectively,
has the Schwarzschild-(anti-)de Sitter solution with
\be
q=\frac{\Mpl}{2}\sqrt{\frac{1}{\eta} \left(\frac{\eta}{\beta}+\Lambda_{\rm b}\right)}, \quad
\Lambda=-\frac{\eta}{\beta},
\ee
which is a self-tuned solution,
since $\Lambda$ is independent of the bare cosmological constant $\Lambda_{\rm b}$.
The sign of $\Lambda$ depends on $\eta$ and $\beta$.
A generalization of this solution to the theory with $G_2=G_2(X), G_4=G_4(X)$ was considered in \cite{Kobayashi:2014eva}. 
These classes also fall into Cases 1-$\Lambda$ and 2-$\Lambda$ for the shift-symmetric Horndeski, which are defined by the conditions~\eqref{case-1L-G} and \eqref{case-2L-G} after plugging \eqref{Horn}.
Note that in the model \eqref{b13} $c_t\neq c$.

The stability of hairy black holes
in the shift-symmetric Horndeski and GLPV theories 
with the scalar field profile $\phi(t,r)=qt+\psi(r)$ has been discussed
in the literature \cite{Ogawa:2015pea,Takahashi:2015pad,Takahashi:2016dnv,Maselli:2016gxk,Babichev:2017lmw,Babichev:2018uiw}.
The odd-parity perturbations about the Schwarzschild[-(anti-)de Sitter] solutions
in the Horndeski theories
were analyzed in Refs.~\cite{Ogawa:2015pea,Takahashi:2015pad,Takahashi:2016dnv,Maselli:2016gxk},
which argued that these solutions suffer either a ghost or gradient instability
in the vicinity of the black hole event horizon.
A possible caveat to the above argument
in terms of the coordinate dependence and boundedness of the Hamiltonian
and the causal properties of perturbations
was presented in Refs.~\cite{Babichev:2017lmw,Babichev:2018uiw}.
The even-parity perturbations and the mode stability about these solutions have not been analyzed yet.
Thus, the stability of these solutions still remains unclear and an open issue.
The same issue will also be the subject for the Schwarzschild-(anti-)de Sitter solutions obtained in this paper,
which will be left for our future study.

A self-tuned solution in the shift-symmetric GLPV subclass with $c_t=c$ was also considered with
\be \label{b17}
G_2(X) = -\Mpl^2\Lambda_{\rm b} + M^6 (-X)^{-1/2}, \quad
G_4(X) = \f{\Mpl^2}{2} + 2M^4(-X)^{-1/2}, \quad
f_4(X) = 2M^4(-X)^{-5/2}, 
\ee
in Eqs.~(25)--(27) in \cite{Babichev:2017lmw}.\footnote{Here, \eqref{b17} and \eqref{b17-s} differ from the expressions in \cite{Babichev:2017lmw} by a factor $M$ as the scalar field is dimensionless in their notation.}
This model indeed satisfies the condition~\eqref{case-1L-G-ctc} for Case 1-$\Lambda$ if
\be \label{b17-s} q = \f{8 M^4}{\Mpl^2 (4 \Lambda_{\rm b} - M^2)}, \quad 
\Lambda = \f{M^2}{4}. \ee
Note that only a positive cosmological constant is allowed in this case.

We find a generalization to the shift-symmetric quadratic DHOST theories with $c_t=c$ is possible by taking
\be \label{ex2}
F_0 = -\Mpl^2 \Lambda_{\rm b} + M^4 h(X), \quad
F_2 = \f{\Mpl^2}{2} + \f{\alpha}{2} M^2 h(X), \quad
A_3 = -8 \beta M^2 \f{h'(X)}{X},
\ee
and other $A_I$'s are determined by the condition~\eqref{ctc}, namely,
\be 
A_1=A_2=0, \quad
A_4= \f{ M^3 h' [ 8 \beta (\Mpl^2 + M^2 \alpha h) 
+  M^2 X (\alpha - 4 \beta) (3 \alpha + 4 \beta) h' ]}{X (\Mpl^2 + \alpha M^2 h)}, \quad 
A_5= - \f{16 M^4 \beta (\alpha - 4 \beta) h'^2}{X (\Mpl^2 + \alpha M^2 h)}, \ee
whereas $F_1(X)$ is an arbitrary function so long as it satisfies the condition $F_{1X}(X_0) = 0$ of Cases 1-$\Lambda$ and 2-$\Lambda$.
Here, $\alpha,\beta$ are a dimensionless model parameter, and $h(X)$ is a function of $X$. 
If $\alpha=4\beta$, the model~\eqref{ex2} falls into the shift-symmetric GLPV theory with $c_t=c$ as the condition \eqref{GLPV} is satisfied, 
in which the model~\eqref{b17} is included as a special case with $h(X)=M^2(-X)^{-1/2}$.
For general $\alpha,\beta$, the model~\eqref{ex2} does not belong to GLPV theory.
Note that the additional condition~\eqref{dec} to prevent a rapid graviton decay can be satisfied by imposing $\beta=0$.

Interestingly enough, as we shall show below, the model~\eqref{ex2} can satisfy all the conditions~\eqref{case-1L-ctc}--\eqref{case-1Lc-ctc} and allow self-tuned/untuned Schwarzschild-(anti-)de Sitter solution with a linearly time dependent profile or constant profile of the scalar field.
First, this model satisfies the condition~\eqref{case-1L-ctc} for Case 1-$\Lambda$ if
\be \label{ex2-s} h(-q^2) = \f{\Mpl^2 [ M^2 + 2 (\alpha - 6 \beta) \Lambda_{\rm b} ]}{M^4 (\alpha - 12 \beta)}, \quad 
\Lambda = - \f{M^2}{2 (\alpha - 6 \beta)} , \ee
where we assume $\alpha\ne 6\beta,12\beta$.
Both the Schwarzschild-de Sitter solution and Schwarzschild-anti-de Sitter solution with positive and negative $\Lambda$ are allowed depending on the sign of $\alpha - 6 \beta$, which is still possible for the GLPV case with $\alpha=4\beta$. 
This solution is a self-tuned Schwarzschild-(anti-)de Sitter solution in the shift-symmetric quadratic DHOST theories with $c_t=c$,
since the cosmological constant $\Lambda$ is determined by the model parameter $M$, which does not depend on $\Lambda_{\rm b}$.
Therefore, if we choose the model parameters $\alpha, \beta, M$ that satisfy $\alpha<6\beta$ and $M\simeq H_0\sqrt{2(6\beta-\alpha)}$, where $H_0$ is the Hubble parameter at the present time, the solution~\eqref{ex2-s} can describe the late-time cosmic acceleration regardless of the bare value $\Lambda_{\rm b}$ of the cosmological constant.
On the other hand, the condition~\eqref{case-2L-ctc} for 
Case 2-$\Lambda$ gives $\Lambda=\Lambda_{\rm b}$ and $h'(X_0)=0$,
which is un-tuned Schwarzschild-(anti-)de Sitter solution.
The model~\eqref{ex2} can also satisfy the condition~\eqref{case-1Lc-ctc} for Case-1-$\Lambda$-c,
so long as the coupling functions are regular at $X=0$, e.g.\ by taking $h(X)\propto X^2$, 
allowing Schwarzschild-(anti-)de Sitter solution with constant scalar profile.
These solutions also exist for $\beta=0$, in which a rapid graviton decay is forbidden.

\section{Conclusion}
\label{sec6}

We clarified the conditions for the shift-symmetric subclass of the quadratic DHOST theories to allow several types of exact black hole solutions, namely, the Schwarzschild and Schwarzschild-(anti-)de Sitter solutions, and the Schwarzschild-type solution with a deficit solid angle, assuming that the scalar field has nontrivial profile with a linear time dependence and that the ordinary kinetic term of the scalar field is constant.
A special case of the scalar field profile includes the static profile with radial dependence only, and the constant profile.
Our analysis also applies to the shift-symmetric subclass such as Horndeski and GLPV theories, and known stealth Schwarzschild solutions and self-tuned Schwarzschild-de Sitter solutions in these theories fall into a special case of our analysis.

Furthermore, we investigated the compatibility between the degeneracy conditions and the conditions to allow the exact solutions obtained in this paper. 
For the Schwarzschild-type solution with a deficit solid angle, we found that the solid deficit angle automatically vanishes in the quadratic DHOST theories with $c_{t}=c$.
For the Schwarzschild and Schwarzschild-(anti-)de Sitter solutions, while
Class I and Class III can allow the solutions, we found a no-go result for Class II
with linearly time dependent profile with $q\ne 0$ and/or radial profile with $\psi'\ne 0$,
so long as we focus on the ansatz of the scalar field given by Eq.~\eqref{phi-qt}.
Namely, for the shift-symmetric subclass of Class II, the Schwarzschild or Schwarzschild-(anti-)de Sitter solution 
exists only for a constant scalar field.
We also provided general DHOST models that satisfy the conditions in the form of Taylor series, as well as specific DHOST and GLPV models that have the novel exact solutions.
It is intriguing to investigate the stability of these solutions, which we leave as a future work~\cite{Takahashi:2019oxz}.

\vspace{5mm}

\noindent
{\bf Note added:} 
The comparison between the results in the present paper and those in \cite{BenAchour:2018dap} in the footnote \ref{note1} is based on the arXiv version 3 of \cite{BenAchour:2018dap}.  
In arXiv version 4 of \cite{BenAchour:2018dap}, the authors corrected 
the corresponding equations.

\acknowledgments{
We thank Jibril Ben Achour and Hongguang Liu for useful discussions.
H.M.\ was supported by JSPS KEKENHI Grant Nos.\ JP17H06359 and JP18K13565.
M.M.\ was supported by FCT-Portugal through Grant No. SFRH/BPD/88299/2012
and the research grant under ``Norma Transit\'oria do DL 57/2016''.
}

\bibliography{ref-BH}

\end{document}